\def\year{2024}\relax
\documentclass[letterpaper]{article} 
\usepackage{dblfloatfix}
\usepackage{booktabs}
\usepackage{aaai22}  
\usepackage{times}  
\usepackage{helvet}  
\usepackage{courier}  
\usepackage[hyphens]{url}  
\usepackage{graphicx} 
\usepackage{natbib}  
\usepackage{caption} 
\DeclareCaptionStyle{ruled}{labelfont=normalfont,labelsep=colon,strut=off} 
\usepackage{algorithm}
\usepackage{algorithmic}
\usepackage{url}

\usepackage{newfloat}
\usepackage{listings}
\floatstyle{ruled}
\newfloat{listing}{tb}{lst}{}
\floatname{listing}{Listing}
\title{Leveraging Recommender Systems to Reduce Content Gaps on Peer Production Platforms}
\author {
    Mo Houtti\textsuperscript{\rm 1}, Isaac Johnson\textsuperscript{\rm 2}, Morten Warncke-Wang\textsuperscript{\rm 2}, Loren Terveen\textsuperscript{\rm 1}
}
\affiliations {
    \textsuperscript{\rm 1} Department of Computer Science \& Engineering, University of Minnesota\\
    \textsuperscript{\rm 2} Wikimedia Foundation\\
    houtt001@umn.edu, isaac@wikimedia.org, mwang@wikimedia.org, terveen@umn.edu
}

\newcommand{\controltabstudyone}{
\begin{table}[]
\centering
\begin{tabular}{@{}l@{}r@{}r@{}}
\toprule
\textbf{Fixed Effect} & \textbf{Regression Coefficient (\textit{t}-statistic)} & \textbf{\textit{p}}\\ \midrule
(Intercept) & -3.57 (-16.61) & *** \\
task type: Cleanup & -0.31 (-3.13) & ** \\
task type: Expand & -0.39 (-3.88) & *** \\
task type: Merge & -1.21 (-8.87) & *** \\
task type: Orphan & -0.35 (-3.34) & *** \\
task type: Stub & 0.20 (2.37) & ** \\
task type: Unencyc. & -0.64 (-5.84) & *** \\
task type: Wikify & -0.65 (-5.86) & *** \\
account age: Q2 & -0.27 (-0.93) & \\
account age: Q3 & -0.89 (-3.11) & ** \\
account age: Q4 & -1.32 (-4.70) & *** \\
pred. class: C/B & 0.24 (3.37) & *** \\
pred. class: A/GA & 0.28 (2.48) & * \\
pred. class: FA & 0.59 (3.49) & *** \\ \bottomrule
\end{tabular}
\caption{Regression coefficients for the control model. Includes a random effect for editor.}
\label{tab:contoltabstudyone}
\end{table}
}

\newcommand{\treattabstudyone}{
\begin{table}[]
\centering
\begin{tabular}{@{}l@{}r@{}r@{}}
\toprule
\textbf{Fixed Effect} & \textbf{Reg. Coefficient (\textit{t}-statistic)} & \textbf{\textit{p}}\\ \midrule
gender: Female & 0.39 (3.31) & *** \\
gender: Non-binary & -12.48 (-0.04) & \\
gender: Not Bio. & -0.19 (-2.57) & * \\ \midrule
geo: Global South & -0.12 (-1.33) & \\
geo: Region-neutral & -0.15 (-2.50) & * \\ \midrule
important topics: True & -0.13 (-1.23) & \\ \bottomrule
\end{tabular}
\caption{Regression coefficients for gender, geography, and important topics models. Each of the three models contains a random effect for editor, the control model's fixed effects, and a fixed effect for the corresponding feature of interest. Features are unordered factors, so there is a separate coefficient for each level of each feature.}
\label{tab:treattabstudyone}
\end{table}
}

\newcommand{\tabstudytwo}{
\begin{table*}[!t]
\centering
\begin{tabular}{@{}llr@{}}
\toprule
\textbf{Fixed Effect} & \textbf{Alternative} & \textbf{Regression Coefficient (\textit{t}-statistic)} \\ \midrule
Gender Treatment & Baseline/Unchanged & -0.12 (-0.80) \\
Gender Treatment & Gender Control & 0.27 (1.37) \\ \midrule
Geography Treatment & Baseline/Unchanged & -0.12 (-0.72) \\
Geography Treatment & Geography Control & 0.02 (0.07) \\ \midrule
Important Topics Treatment & Baseline/Unchanged & -0.13 (-1.03) \\
Important Topics Treatment & Important Topics Control & 0.33 (1.37) \\ \bottomrule
\end{tabular}
\caption{Regression coefficients for each of the 6 models in Study 2.}
\label{tab:tabstudytwo}
\end{table*}
}

\newcommand{\editedrecs}{
\begin{figure}[!b]
    \centering
    \includegraphics[width=.45\textwidth]{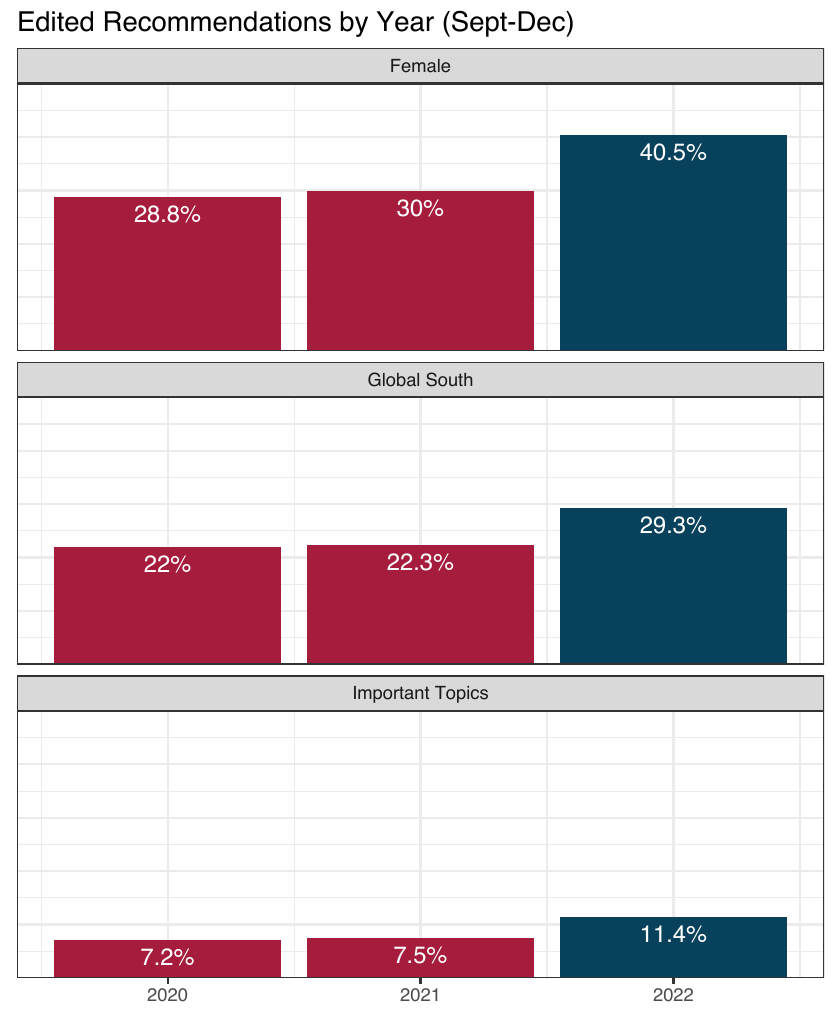}
    \caption{Percentage of edited recommendations that belonged to underrepresented categories. Previous years are in red and study period is in blue. Graphs show consistent increases across all categories, likely due to our intervention.}
    \label{fig:edited_recs}
\end{figure}
}

\newcommand{\samplerecs}{
\begin{figure}[!t]
    \centering
    \includegraphics[width=.45\textwidth]{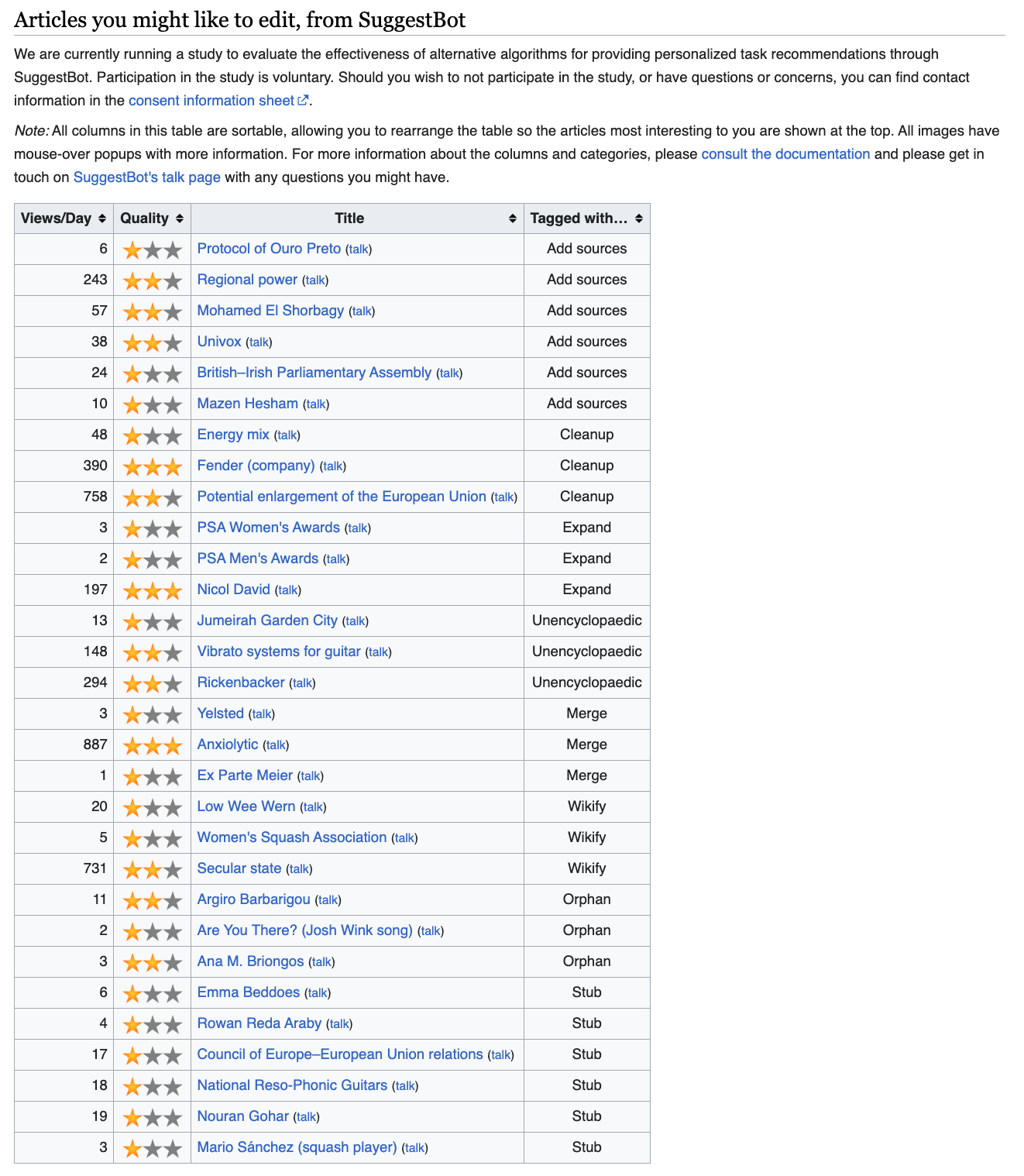}
    \caption{An example set of recommendations, posted by SuggestBot to a user's Talk page. Roughly half of the recommendations are part of an experimental group, but the user is not given any indication as to which.}
    \label{fig:samplerecs}
\end{figure}
}

\newcommand{\filterchart}{
\begin{figure}[!t]
    \centering
    \includegraphics[width=.3\textwidth]{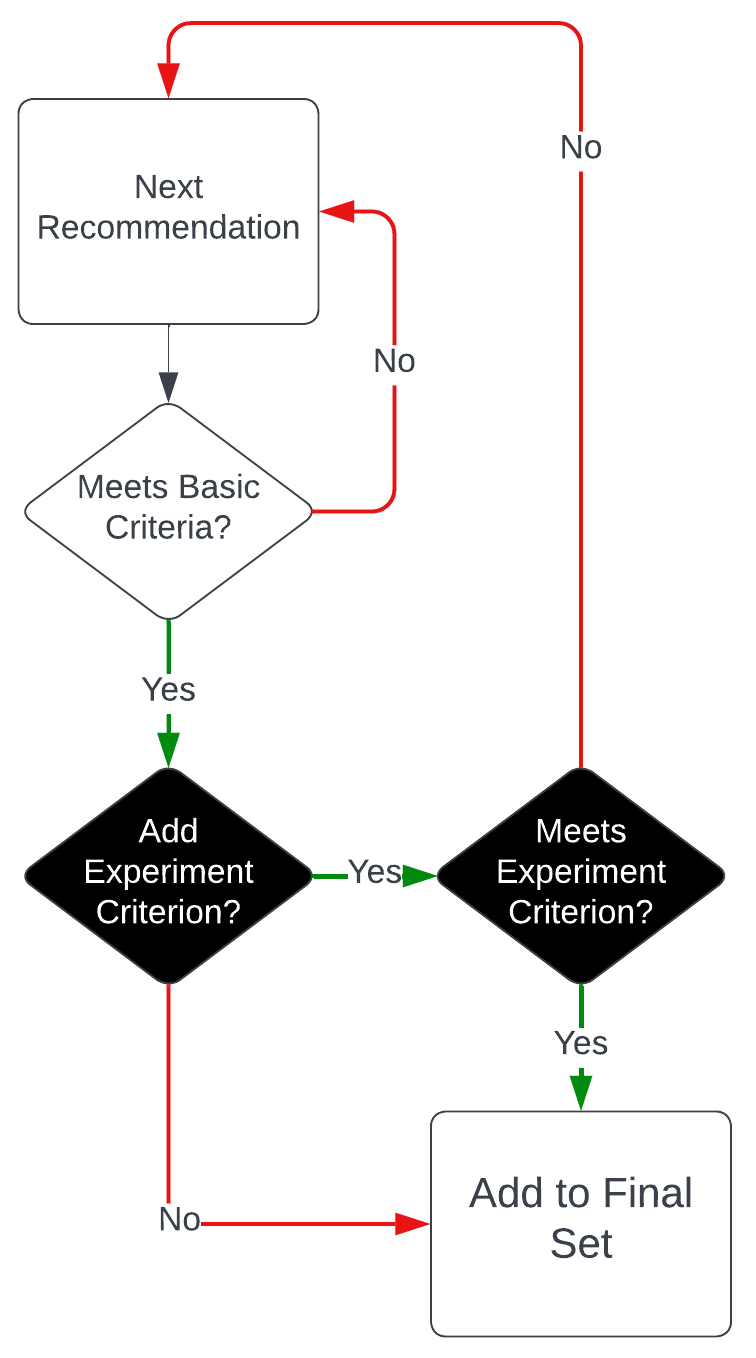}
    \caption{Diagram describing SuggestBot's process of filtering recommendations, and how we intervened in the filtering process to produce treatment groups. Black elements represent additions for the experiment. ``Add Experiment Criterion'' was given a random 55\% chance of being yes.}
    \label{fig:filter}
\end{figure}
}

\urldef\myurl\url{https://en.wikipedia.org/wiki/User_talk:SuggestBot#Is_there_a_way_to_increase_%20temperature%20}

\begin{document}

\maketitle

\begin{abstract}
Peer production platforms like Wikipedia commonly suffer from content gaps. Prior research suggests recommender systems can help solve this problem, by guiding editors towards underrepresented topics. However, it remains unclear whether this approach would result in less relevant recommendations, leading to reduced overall engagement with recommended items. To answer this question, we first conducted offline analyses (Study 1) on SuggestBot, a task-routing recommender system for Wikipedia, then did a three-month controlled experiment (Study 2). Our results show that presenting users with articles from underrepresented topics increased the proportion of work done on those articles without significantly reducing overall recommendation uptake. We discuss the implications of our results, including how ignoring the article discovery process can artificially narrow recommendations on peer production platforms. 
\end{abstract}

\section{Introduction}

Wikipedia’s gender, geographical, and other content disparities have long been documented and criticized in the popular media~\cite{baltz_wikipedias_nodate, cohen_define_2011, gleick_wikipedias_nodate, mandiberg_mapping_2020, resnick_2018_2018, torres_why_2016} and research literature~\cite{antin_gender_2011, beytia_positioning_2020, bjork-james_new_2021, hargittai_mind_2015}. Both the Wikimedia Foundation (non-profit that supports Wikipedia) and editor-led communities have devoted substantial attention to addressing the problem over the past several years~\cite{halfaker2017interpolating,langrock2022gender,scott_why_2018, noauthor_strategywikimedia_2017, noauthor_open_2023} but, while these efforts have succeeded in reducing some gaps, there remains much to be done. For example, one of the most easily quantifiable and widely acknowledged disparities on Wikipedia is the gender gap---yet as of April 2023, biographies about women still only represent 19.5\% of all biographies on English Wikipedia~\cite{wikipedia_wikipediawikiproject_2023}.

We propose leveraging recommender systems to reduce these content disparities. Task-routing recommender systems have already been deployed successfully in Wikipedia~\cite{cosley_suggestbot_2007, wulczyn2016growing}, but their algorithms tend to optimize for editor interest or predicted reader need alone. By modifying them to additionally consider \textit{content equity}---i.e. prioritizing underrepresented topics---these systems could guide editors towards work that would help to reduce content disparities on Wikipedia.

However, editor interest-based recommender systems have been shown to increase editing by \textit{four times} compared to recommending random articles~\cite{cosley_suggestbot_2007}, and re-orienting them in the direction of content equity could reduce these benefits. Indeed,~\citeauthor{warncke-wang_misalignment_2015} (\citeyear{warncke-wang_misalignment_2015}) caution against ``simplistic attempts'' to push editors towards work they are not interested in, lest they edit less or leave the platform altogether. We agree there is good reason for caution, but other work also gives us reason to be optimistic. Through qualitative analysis of editors' discussions,~\citeauthor{houtti_we_2022} (\citeyear{houtti_we_2022}) showed that editors consider content balance---including along gender and geographical lines---to be an essential factor in deciding how articles should be prioritized for improvement. Yet those same editors compiled lists of high-priority articles that were \textit{not} balanced along those dimensions. Based on this,~\citeauthor{houtti_we_2022} speculate that editors are at least \textit{willing} to prioritize articles from under-represented categories, but that self-focus bias~\cite{hecht_measuring_2009} leads them to more readily identify articles salient to their own experiences instead:

\begin{quote}
``On one hand, I'm surprised it [Menstruation article] isn't here, but then as one of the x-deficient 90\% of editors, I wouldn't have even thought to add it.''
\end{quote}

Similarly, many editors might neglect articles from underrepresented topics not because they are averse to editing those articles, but because they more readily identify articles salient to their own experiences---ones that are more male and more western, among other things~\cite{houtti_we_2022, warncke-wang_misalignment_2015}.

If this is true, \textbf{content disparities could be reduced by simply making underrepresented articles more visible to editors}. Indeed, doing so might end up better aligning with editor values that are simply not reflected in the edit histories that task-routing algorithms rely on. This alignment could realistically lead to \textit{more} editing, as~\citeauthor{nov_what_2007} (\citeyear{nov_what_2007}) found that ideology and values are strong motivators for Wikipedians to contribute.

While this does make sense in theory, whether guiding editors towards under-represented articles would \textit{actually} cause a meaningful increase or decrease in editing is an empirical question. We therefore conducted two empirical studies on SuggestBot---a task-routing recommender system for Wikipedia~\cite{cosley_suggestbot_2007}. We first analyzed articles recommended by SuggestBot in 2021 (Study 1) and found, among other things, that editors were \textit{more} likely to edit biographies of women than biographies of men. We then conducted a three-month controlled experiment on SuggestBot, where we replaced a subset of editors' recommendations with the most relevant articles from underrepresented categories (Study 2). We found that these alternative recommendations did not suffer from any significant decreases in uptake. Moreover, providing a higher number of recommendations from underrepresented categories substantially increased the share of recommendation-prompted editing on articles from those categories. Our paper contributes empirical findings that support the use of recommender systems to help reduce content disparities on Wikipedia and other peer-production platforms. In particular, we emphasize how edit history-based inference fails to acknowledge that editing behavior is largely determined by what content the editor discovers in the first place---an oversight that can lead systems to needlessly magnify self-focus bias. We discuss how this may generalize to other platforms that employ recommender systems driven by behavioral data.

We begin by covering the prior research in this area, provide an overview of the SuggestBot system, outline the methods and results of each study, and conclude with a discussion of our findings' implications.

\section{Related Work}

Below we briefly outline the relevant tensions related to content gaps on Wikipedia (the context), the role of recommender systems on Wikipedia (our focus), and research about aligning recommender systems and goals related to fairness or equity (our tested intervention).

\subsection{Content Gaps on Wikipedia}
Content gaps on Wikipedia are well-documented (see~\citeauthor{redi_taxonomy_2021} \citeyear{redi_taxonomy_2021} for a recent review) and while much absolute progress has been made---e.g., there are many more (high-quality) articles about women than ten years ago thanks to organized efforts to close these gaps~\cite{halfaker2017interpolating,langrock2022gender}---the distribution of content on Wikipedia continues to hold major representational biases.\footnote{For example, see \url{https://humaniki.wmcloud.org/} for statistics about the gender gap.} While these gaps are the result of many complex processes, generally it is understood that a major contributing factor is that of self-focus bias~\cite{hecht_measuring_2009}. Self-focus bias is the concept that Wikipedians edit about content that is familiar and of interest to themselves, so a community of editors that are not representative of the world's population~\cite{2021communityinsights} will not produce an encyclopedia that is representative of the world's knowledge.

The Wikimedia community has responded to the issue of content gaps through initiatives such as adopting a universal code of conduct~\cite{ribewikimedia} to address harassment issues known to be a major barrier to gender equity~\cite{genderequity2018}, and building partnerships with publishers to provide editors with free access to reliable sources that they can incorporate to reduce barriers in accessibility of digital sources.\footnote{\url{https://en.wikipedia.org/wiki/Wikipedia:The_Wikipedia_Library}} Perhaps the most direct and visible approach has been through organizing campaigns to close these gaps---e.g., by creating biographies of women~\cite{langrock2022gender,tripodi2021ms}, improving content on important topics~\cite{houtti_we_2022}, or contributing imagery of cultural heritage from around the world~\cite{azizifard2022wiki}. These campaigns help to attract and socialize newcomers~\cite{farzan2016bring} as well as focus attention of existing editors on closing these gaps. This collective-action approach has been taken by other peer-production communities such as OpenStreetMap, which for example has organized extensive humanitarian mapping  initiatives~\cite{herfort2021evolution} to help address their geographic content gaps~\cite{thebault2018geographic}.

Self-focus bias would indicate that in the long-term, Wikipedia and other peer-production communities need a more representative community of editors in order to close content gaps. In the meantime, these campaigns represent an important effort to make content more representative. There are open questions, however, about how to support the current Wikipedia community in overcoming some of this self-focus bias to better address these gaps.

\subsection{Recommender Systems on Wikipedia}
A separate set of initiatives largely aimed at making Wikipedia easier to edit (and therefore reducing the barriers to participation) has focused on building recommender systems for Wikipedia editors. These have included tools like personalized recommendations of articles to edit~\cite{cosley_suggestbot_2007} or translate ~\cite{wulczyn2016growing} based on edit history and reader needs, and structured tasks to assist newcomers in learning how to edit Wikipedia~\cite{gerlach2021multilingual}.

These recommender systems (and other sociotechnical tools aimed at supporting newcomers such as The Wikipedia Adventure~\cite{narayan2017wikipedia} and Wikipedia Teahouse~\cite{morgan2018evaluating}) have largely been evaluated on their ability to increase edit counts or improve editor retention. While they may indirectly help close content gaps by reducing participation barriers, there has been less attention to how they might be more explicitly aligned with campaigns' efforts to close content gaps. 

\subsection{Fairness in Recommender Systems}
Within the recommender system and information retrieval literature, research has begun to examine how adjustments to standard algorithms can help to incorporate values such as fairness. There are many complexities to this work~\cite{sonboli2022multisided} but of particular interest to Wikipedia are systems that seek to balance user personalization with fairness with respect to the items being recommended---i.e. helping editors find topics relevant to their interests, while also incorporating preferences for the overall distribution of items recommended across all the users, such as \citeauthor{mehrotra2018towards}~\citeyear{mehrotra2018towards} and \citeauthor{sonboli_opportunistic_2020}~\citeyear{sonboli_opportunistic_2020}. These systems suggest a means by which self-focus bias can be tailored to better align with the goals of campaigns. Most systems have been evaluated only in offline settings and not in a complex in-the-field setting like Wikipedia, in which a recommendation to improve an article is likely a much larger ask than suggesting a song or other piece of media to consume.

\samplerecs

\section{Background: SuggestBot}

SuggestBot is a recommender system on Wikipedia that recommends work tasks based on editor interest, as inferred from past edit history (see~\citeauthor{cosley_suggestbot_2007}~\citeyear{cosley_suggestbot_2007} for details). Editors can request a single set of recommendations, or subscribe to receive personalized recommendations at a configurable interval---e.g., every 2 weeks. In either case, SuggestBot provides a set of 30 articles (Figure~\ref{fig:samplerecs}), each annotated with metadata about the article's quality and popularity as well as an indication of the kind of work needing to be done on the article---e.g., ``Add Sources''.\footnote{These task types are manually assigned to articles by other Wikipedia editors in the course of their editing.}

\section{Methods (Study 1)}
Study 1 was an offline analysis of the efficacy of SuggestBot recommendations. The goal was to establish a baseline understanding of how editing behavior is impacted by factors relevant to content gaps and inform the subsequent experiment (Study 2).

\subsection{Dataset and Features}
The period for Study 1 was 2021 in its entirety. We collected all SuggestBot recommendations during this period. To ensure we had adequate data for each editor, we removed any editors who made fewer than 100 edits during the study period from our dataset. We then removed any recommendations made while the editor was inactive. We considered an editor inactive if they made no edits to \textit{any} articles in the 30-day period following the recommendation. Our final dataset contained 82,650 recommendations (2755 sets of 30) across 375 users.

For our outcome variable, each recommendation was labeled as successful (binary) if the editor made 1 or more edits to the recommended article within 30 days of receiving the recommendation.

We next identified relevant article characteristics that were likely to impact editing behavior, to be used as control variables:

\begin{itemize}
\item{\textbf{views}: the number of article views, by quartile within the dataset. (Quartiles 1-4)}
\item{\textbf{top importance rating}: the highest importance rating assigned to the article by a WikiProject. (Unrated, Low, Mid, High, Top)}
\item{\textbf{article year}: the period to which the article most closely pertains. (Unknown, Pre-1900s, 20th century, 21st century)}
\item{\textbf{predicted class}: the article's quality class, as predicted by the ORES system. (Stub/Start, C/B, A/GA, FA)}
\item{\textbf{assessed class}: the article's quality class, as manually assessed and tagged by Wikipedia editors. (None, Stub/Start, C/B, A/GA, FA)}
\item{\textbf{task type}: the kind of work needing to be done, as indicated by SuggestBot. (Add sources, Cleanup, Expand, Unencyclopaedic, Merge, Wikify, Orphan, Stub)}
\end{itemize}

We also included editor characteristics likely to affect editing behavior as control variables:

\begin{itemize}
\item{\textbf{account age}: the age of the editor's account, by quartile within the dataset. (Quartiles 1-4)}
\item{\textbf{total edit count}: the editor's total number of edits on English Wikipedia, by quartile within the dataset. (Quartiles 1-4)}
\end{itemize}

Finally, we included features to represent various dimensions of content equity: 

\begin{itemize}
\item{\textbf{gender}: the gender of the article's subject, if it is a biography. (Not Biography, Male, Female, Other)}
\item{\textbf{geography}: the region to which the article pertains (Global North, Global South, Both/Region-Neutral)}
\item{\textbf{important topics}: whether the article is included in a WikiProject on an important topic.}
\end{itemize}

We selected gender and geography because they are easy to operationalize and are two of the most commonly cited dimensions along which English Wikipedia is known to have content disparities. We gathered gender\footnote{Specifically, for articles about humans with the sex-or-gender property, we separated articles into (cisgender) Male, (cisgender) Female, and a final category that incorporated transgender and non-binary gender identities. Given the low proportion of non-binary gender identities represented on Wikipedia, we grouped them with the Female group for our experiment.} and coordinate information on the articles in our dataset from Wikidata. 

Our inclusion of important topics was motivated by its inclusion in \citeauthor{redi_taxonomy_2021}'s (\citeyear{redi_taxonomy_2021}) taxonomy of knowledge gaps on Wikimedia projects. Though the boundaries of what should be considered an important topic have not been established by the Wikimedia community, we chose to operationalize it as topics directly relevant to the United Nation's Sustainable Development Goals,\footnote{\url{https://meta.wikimedia.org/wiki/Movement_Strategy/Recommendations/Identify_Topics_for_Impact#What}} which worked out to 11 WikiProjects: Disability, Politics, Agriculture, Medicine, Education, Water, Sanitation, Energy, Environment, Climate change, and Human rights.

\subsection{Analysis}
We fit generalized linear mixed models (GLMMs) with a binomial distribution to identify relationships between our variables of interest and our binary outcome variable. All features were encoded as categorical (unordered factors).

We first constructed a model containing only control variables. We started with the simplest model, which included a random effect for editor and no fixed effects. We iterated through each of our control variables, fitting a new GLMM that included that feature as a fixed effect. If any of the new features significantly improved fit (as determined by $\chi^2$ likelihood-ratio tests), we adopted the model that was most improved. In essence, we iteratively added the control variable that most improved model fit, then repeated this process until none of the remaining control variables significantly improved fit when added as fixed effects.

Then, for each of our variables of interest, we fit a new GLMM, adding the variable of interest as a fixed effect alongside the selected control variables. We again used $\chi^2$ likelihood-ratio tests to compare the fit of each of these models to the control model.

\section{Results (Study 1)}

\controltabstudyone

\subsection{Control Model}

The control model included three fixed effects---task type, account age, and predicted class. All other potential control features did not significantly improve the model, so we did not include them. We report the regression coefficients for this model in Table~\ref{tab:contoltabstudyone}. Our control model shows that higher quality articles are edited more than lower quality articles, and that editors who have been registered for longer are less likely to edit recommended articles. It also shows that articles labeled with easier tasks (e.g., Add Sources) are edited more frequently than those labeled with harder tasks (e.g., Wikify). It is worth noting here that article merges are not recorded as edits, which explains the negative coefficient for that task type. We kept recommendations with the Merge task type in our analysis anyway, however, because editors may still work on recommended articles in ways that do not fit with the task type presented by SuggestBot.

\subsection{Gender}

Adding a fixed feature for gender significantly improved the model ($p < 0.001$). The model showed a positive regression coefficient (Table~\ref{tab:treattabstudyone}) for Female recommendations ($p < 0.001$), with Male in the intercept, indicating a higher likelihood of a Female article recommendation leading to an edit. The coefficient for Non-binary recommendations was not significant due to low sample size.

\subsection{Geography}

Adding a fixed effect for geography, once again, significantly improved model fit ($p = 0.042$). Here, however, the better fit seems to be driven by differences between Region-Neutral and non-Region-Neutral articles; with Global North in the intercept, the coefficient for Global South is not significant ($p = 0.182$), and the coefficient for Both/Region-Neutral is significant ($p = 0.013$).

\subsection{Important Topics}

Adding a fixed feature for important topics did not significantly improve model fit ($p = 0.211$). The coefficient for important topics is slightly negative but not significant ($p = 0.221$).

\treattabstudyone

\subsection{Observational Study to Controlled Experiment}

Study 1 gave us a baseline understanding of how Wikipedians react to SuggestBot's recommendations. Overall, it paints an encouraging picture of editors' reactions to articles from underrepresented categories. However, observational studies are inherently limited. We may not have incorporated all possible control variables, and doing so would likely make our models unmanageably complex anyway. Moreover, even articles from underrepresented categories were still highly relevant to the editor---otherwise the standard SuggestBot algorithm would not have included them in the first place. For a recommender system to truly have impact, we would need to intentionally present editors with items deemed less relevant. Unfortunately, an observational study does not allow us to generalize to this new paradigm, where content equity has to come at some cost to item relevance. This led us to conduct a controlled experiment in Study 2.

\section{Methods (Study 2)}

\subsection{Experimental Design}

\filterchart

SuggestBot is an ensemble recommender system; it begins by generating three large (1000+ items) sets of candidate recommendations using three different algorithms. It then iterates through each set---starting with the most relevant item in the set and going in descending order---and selects the first item from the set that meets its basic filtering criteria (e.g., has not already been recommended to this user recently). It cycles through the 3 candidate sets, selecting one recommendation from each, until it has assembled 30 recommendations which are then served to the user.

Even though the recommendations are sorted in relevance order, SuggestBot must go through a lengthy filtering process because it must also select recommendations that are tagged with the appropriate task type. For example, SuggestBot always includes 6 ``Add Sources'' tasks in a recommendation set; it achieves this by adopting ``is tagged with Add Sources'' as a filtering criterion in 6 of its iterations. This means that, on average, SuggestBot filters through 461 recommendations before finding a suitable recommendation that meets its basic criteria.

We implemented our experiment by intervening in the filtering process, after the initial candidate recommendation sets were generated (Figure~\ref{fig:filter}). More specifically, we modified SuggestBot's logic to, in a random 55\% of iterations, adopt an additional filtering criterion corresponding to one of our experimental groups. It would only select a candidate article to be recommended if the article met all the original filtering criteria \textit{and} the new criterion. For our gender experimental group, for example, SuggestBot adopted the filtering criterion ``is a biography about a woman''---i.e., it selected the most relevant article from the candidate set \textit{that was a biography about a woman}.

As previously mentioned, adding new filtering criteria meant our treatment articles would, in the aggregate, have lower relevance scores than our unchanged articles. To isolate the effects of these relevance changes from the effects of changes in the article features we were concerned with, we also generated lower-relevance control groups. This would allow us to answer questions like \textit{``did editors react differently to articles about women because they were less relevant, or because they were articles about women?''}

Therefore, if a recommendation was set to receive an additional filtering criterion, it was also given a 50\% chance of being a lower-relevance control. In this case, SuggestBot would find the first article meeting all the original filtering criteria and the experimental criterion, but then select the first article \textit{after it} that met all the original filtering criteria but did \textit{not} meet the added experimental criterion. Using gender as an example, the end result was that we created two groups of recommendations, both of which suffered similar relevance drops, one of which was composed entirely of articles about women (treatment), while the other contained no articles about women (reduced relevance control).

Overall, each recommended article had a probability of belonging to one of the following groups:

\begin{itemize}
    \item {45\%: unchanged/baseline}
    \item {9.17\%: gender treatment (female and non-binary articles)}
    \item {9.17\%: geography treatment (global south articles)}
    \item {9.17\%: important topics treatment (important topics articles)}
    \item {9.17\%: gender control (reduced relevance non-female or non-binary articles)}
    \item {9.17\%: geography control (reduced relevance non-global-north articles)}
    \item {9.17\%: important topics control (reduced relevance non-important-topics articles)}
\end{itemize}

Sometimes, no suitable article could be found for inclusion in an experimental group. In those cases, SuggestBot reverted to selecting the most relevant article using only the basic filtering criteria, and the article was considered to be part of the unchanged/baseline group. Therefore, the actual group breakdown of recommendations was:

\begin{itemize}
    \item {19,805 (49.5\%): unchanged/baseline}
    \item {3,781 (9.5\%): gender treatment (female and non-binary articles)}
    \item {3,284 (8.2\%): geography treatment (global south articles)}
    \item {3,361 (8.4\%): important topics treatment (important topics articles)}
    \item {3,533 (8.8\%): gender control (reduced relevance non-female or non-binary articles)}
    \item {3,181 (8.0\%): geography control (reduced relevance non-global-north articles)}
    \item {3,045 (7.6\%): important topics control (reduced relevance non-important-topics articles)}
\end{itemize}

We added text to each recommendation set's intro message to inform users that an experiment was being run, and allow them to opt out. Only a single user opted out of the experiment; they were given unchanged recommendations and removed from our dataset.

\subsection{Dataset}

The experiment was conducted over approximately 3 months, from September 7th to December 31st, 2022. Our dataset included all SuggestBot recommendations served in this period while the receiving editor was active. We again defined active as having made at least one edit to \textit{any} article in the 30 days following recommendation. The final dataset contained 39,990 recommendations (1,333 sets of 30) across 281 unique users with a recommendation being labeled as successful if the editor subsequently edited the article within the next 30 days.

\subsection{Analysis}

Our experimental intervention could have plausibly reduced engagement across \textit{all recommendations}, even those in the baseline/unchanged group. We therefore first generated descriptive statistics to compare overall recommendation uptake in the context of previous years. We then verified that recommendations in our reduced relevance control groups did in fact have similar relevance to their corresponding treatment group recommendations. Finally, we once again fit generalized linear mixed models (GLMMs) to model the relationships between each of our experiment variables and the same binary outcome as in Study 1---whether the recommendation did or did not prompt work on the article.

\tabstudytwo

For each of our features of interest (gender, geography, and important topics), we fit two GLMMs. One GLMM was fit solely on recommendations that were part of the intervention and contained a fixed effect representing whether a recommendation was in the treatment group or its corresponding lower-relevance control group. The goal of this first model was to examine whether the feature had an effect on editing \textit{assuming relevance were held constant}. The other contained a fixed effect for whether a recommendation was in the treatment group or in the unchanged/baseline group, and was used to examine the effects of replacing SuggestBot's normal recommendations with items from underrepresented categories. Each model was fit using only the recommendations that fit into one of its corresponding groups.

All models included editor as a random effect. For each of the GLMMs, we used a $\chi^2$ likelihood-ratio test to determine whether adding the fixed effect significantly improved model fit.

\editedrecs

\section{Results (Study 2)}

\subsection{Descriptive Statistics}

\subsubsection{Experimental intervention did not reduce overall uptake}

During our study period, SuggestBot recommendations prompted a total of 1090 edits from 599 of the 39,990 recommendations (1.5\% uptake with approximately 2 edits per successful recommendation). Of those 1090 edits, 32\% included changes to the article content, 23\% included changes to internal links, 24\% included changes to references, and 68\% included changes to templates.\footnote{Note that a single edit often included multiple types of changes and the SuggestBot tasks are marked via templates so the high rate of template editing is not surprising in this context.}

By comparison, uptake percentages for the same months in 2020 and 2021 were 1.9\% and 2.0\%, respectively. We suspected this year-over-year reduction was caused by global editing trends on English Wikipedia, rather than our experimental recommendations. For instance, the spikes in edit activity that occurred with the beginning of the COVID-19 pandemic~\cite{ruprechter2021volunteer} have lessened in recent years.\footnote{\url{https://stats.wikimedia.org/#/en.wikipedia.org/contributing/edits/normal|bar|2019-05-06~2023-05-13|editor_type~user+(page_type)~content|monthly}} To confirm, we compared year-over-year recommendation uptake for the month of August (right before our study period), and found 1.9\%, 1.6\%, and 1.2\% for 2020, 2021, and 2022, respectively. The fact that uptake \textit{before} our study period was already lower than previous years---and that it then \textit{increased} during the study period---confirmed that global editing trends were likely responsible for these changes. We therefore proceeded to examine what percentage of these edited recommendations belonged to underrepresented categories.

\subsubsection{More edited recommendations were female}

Of 3,781 recommendations in the gender treatment group, 57 (1.5\%) prompted editing; this was virtually identical to overall uptake. We were primarily concerned with the ratio of female-to-male articles, so only biographies were included in the following calculations. We found that 40.5\% of edited biography recommendations were female during our study period---a substantially higher portion than the 28.8\% and 30.0\% in 2020 and 2021, respectively. In short, presenting editors with a greater share of female recommendations---41.7\% female recommendations, instead of 22.3\% and 23.0\% in 2020 and 2021---also resulted in a greater share of \textit{editing} being done on female biographies as compared with male biographies.

\subsubsection{More edited recommendations pertained to global south}

Of 3,284 recommendations in the geography treatment group, 50 prompted editing (1.5\% uptake); once again, this was very similar to overall uptake. We were primarily concerned with the ratio of global-south-to-north articles, so only non-region-neutral articles were included in the following calculations. We found that 29.3\% of edited recommendations pertained to the global south, compared with 22.0\% in 2020 and 22.3\% in 2021. This was a smaller increase than for gender, likely because our experiment changed the baseline share of global south recommendations by a much smaller amount than it did for female recommendations; the percentages of global south recommendations were 28.2\%, 27.7\%, and 30.9\% in 2020, 2021, and 2022. Presenting marginally more global south recommendations, however, seems to have still resulted in more editing on global south articles.

\subsubsection{More edited recommendations were from important topics}

Of 3,361 recommendations in the important topics treatment group, 37 prompted editing (1.1\%). However, we found that 11.4\% of edited recommendations in our study period pertained to important topics. This was, again, higher than the two previous years---7.2\% in 2020 and 7.5\% in 2021. The percentage of important topics recommendations presented increased from 10.2\% in 2020 and 9.7\% in 2021, to 16.0\% during our 2022 study period. Once again, recommending more important topics articles resulted in a greater share of \textit{edited} recommendations being important topics articles.

\subsubsection{Treatment and reduced relevance control groups had substantial relevance drops}

SuggestBot is an ensemble recommender system, so there was no single scale we could use to meaningfully compare recommendations' relevance. To verify that the reduced relevance control groups had approximately similar relevance to their corresponding treatment groups, we compared the number of candidate recommendations SuggestBot had to filter out before finding a suitable recommendation for each group. Candidate recommendations were evaluated in decreasing order of relevance within each sub-recommender, so fewer filtered-out candidates generally indicated a more relevant recommendation (and vice versa).

Recall that, for baseline/unchanged recommendations, SuggestBot had to filter out 461 candidate articles before finding a suitable recommendation that met its basic criteria. On the other hand, the means for the gender, geography, and important topics treatment groups were 865, 697, and 789 candidates, respectively, indicating substantial drops in relevance as expected. The means for the lower relevance control groups were similar to those of their corresponding treatment groups: 834, 675, and 750, respectively, confirming that lower relevance control groups were roughly equivalent in relevance to their treatment group counterparts.

\subsection{Generalized Linear Mixed Models}

Our descriptive statistics so far tell a straightforward story: presenting SuggestBot users with more underrepresented items increased the share of editing done on underrepresented items. We now present results from our analysis using generalized linear mixed models (GLMMs), which allowed us to introduce a random effect for editor and isolate the effects of the features we were interested in (Table~\ref{tab:tabstudytwo}).

\subsubsection{Gender fixed effects not significant}

Our first GLMM for gender compared female treatment group recommendations with baseline/unchanged recommendations. Adding a fixed effect representing those features did not significantly improve the model over one with just a random effect for editor ($p = 0.42$), indicating that editors were open to editing biographies of women at about the same rate as their normal recommendations. We fit a second GLMM to compare female treatment group recommendations with their reduced-relevance controls and, as expected, also found no significant model improvement ($p = 0.17$).

\subsubsection{Geography fixed effects not significant}

We fit a GLMM to compare global south treatment group recommendations with baseline/unchanged recommendations. Once again, adding a fixed effect for these features did not significantly improve our model ($p = 0.47$), indicating that editors were open to editing articles pertaining to the global south at about the same rate as their normal recommendations. The GLMM comparing global south treatment recommendations with their reduced-relevance controls also was not significantly better than the random-effect-only alternative ($p = 0.94$).

\subsubsection{Important topics fixed effects not significant}

Our results here are similar to those for the gender and geography experimental groups. Adding a fixed effect representing whether an article was in the important topics treatment or baseline/unchanged group did not significantly improve our model ($p = 0.3$). Once again, this shows that editors are largely willing to work on these articles at the same rate. The model comparing important topics treatment recommendations with their reduced-relevance control recommendations was also not significantly better than the control model ($p = 0.17$).

\subsubsection{Summary}

Our ``null'' results are an exciting outcome; editors show a willingness to edit content on underrepresented topics, even when those items are less relevant. While not significant, the regression coefficients comparing each treatment group with baseline show small decreases in predicted uptake, while the ones comparing each treatment group with its lower-relevance control show small increases. This substantiates our findings from Study 1; \textit{relevance being equal}, recommendations from underrepresented categories have a slightly \textit{higher} chance of being edited. When relevance decreases substantially to provide more underrepresented items, we do see reduced uptake, but not by enough to significantly offset the increased editing on underrepresented items. We now discuss the implications of these results.

\section{Discussion}

\subsection{Peer Production Filter Bubbles}

Personalized task-routing recommender systems like SuggestBot aim to provide the recommendations that are most likely to result in edits. Therefore, in theory, using edit history to personalize recommendations makes sense---we can figure out what an editor \textit{would} edit by looking at what they have already edited---and research has borne this out~\cite{cosley_suggestbot_2007,wulczyn2016growing}. However, this ignores that the process by which editors \textit{discover} articles also determines what they are likely to edit. In practice, therefore, recommender systems can falsely infer \textit{editing} preferences from biases in an editor's article discovery process. 

Let us imagine, for example, a male editor who loves editing biographies. As he watches the 2022 FIFA World Cup final, he looks up each player's Wikipedia article and makes edits to a large subset of them. A recommender system might then recommend articles similar to those---perhaps other sports-related articles, or more biographies about men. In reality, however, the editor is perfectly happy to work on \textit{any biographies he encounters}. Despite this, the recommender system reinforces the self-focus bias that crept in through the editor's article discovery process. This may cause editors to receive recommendations that are narrower than they would like, sometimes in ways that are noticeable to them.\footnote{See an example from a SuggestBot user:\\~\myurl
} 

In summary, recommender systems that use edit history but do not incorporate the article discovery process may risk unnecessarily narrowing editors' recommendations. What we identify here is similar to the ``filter bubble''~\cite{pariser_filter_2011} phenomenon, whereby recommender systems may narrow a user's content consumption. Most prior research has studied filter bubbles in the context of news and ideological polarization~\cite{bakshy_exposure_2015,haim_burst_2018,guess_how_2023}, where people undoubtedly have different information-seeking behaviors and different psychological incentives---e.g., greater risk of selective exposure~\cite{dahlgren_critical_2021}---than in peer production. The outcomes of this research are mixed~\cite{bakshy_exposure_2015,bechmann_are_2018,michiels_what_2022}, and we know of no empirical investigations of filter bubbles in our context, but our findings here and prior work on self-focus bias~\cite{hecht_measuring_2009} suggest a filter bubble hypothesis is plausible.

Recommender systems are themselves a means of discovering content, so they can be leveraged to \textit{counteract} self-focus bias in the discovery process. The recommender system literature has explored ways of counteracting various biases over the years, moving from more general concepts of diversity to more targeted fairness objectives~\cite{ekstrand2022fairness}. Indeed, our findings show that this can work; when we presented editors with more articles from underrepresented categories, they \textit{edited} more articles from those categories. If they were averse to editing these articles, we would have expected to see a significant drop in recommendation uptake---yet we did not. This signals that content gaps are not entirely driven by Wikipedians' editing preferences; biases in the article discovery process likely play a significant role. In fact, based on~\citeauthor{houtti_we_2022}'s (\citeyear{houtti_we_2022}) analysis of English Wikipedia's values, it would not be at all surprising if Wikipedians' editing preferences were \textit{in favor} of articles from underrepresented topics. Our Study 1 results support this conclusion for gender and, while not significant, the coefficients in our Study 2 models comparing each treatment group with its lower-relevance control do as well.

\subsection{Incorporating Personal Diversity Tolerance}

In Study 2, we implemented aggressive, consistent treatments to keep the recommender simple while ensuring that effects could be detected if they were of a reasonable size. Our analysis controlled for individual editors through a random effect but did not explore individual editing differences. Prior work provides more sophisticated means of increasing content equity that do take individual differences into account. \citeauthor{sonboli_opportunistic_2020} (\citeyear{sonboli_opportunistic_2020}), for example, describe an algorithm that diversifies recommendations only along dimensions where the user has shown tolerance for diversity. Attempts to leverage recommender systems to reduce content gaps should take advantage of these methods to both minimize the risk of pushing away editors with low diversity tolerance and to maximize the equity benefits gained from editors with high diversity tolerance, even if that tolerance is only along particular dimensions.

\subsection{Recommender Systems Are \textit{Part} of a Solution}

It is clear from our results that recommender systems could play a significant role in reducing content gaps, but can they eliminate gaps altogether? Sadly, no. We see recommender systems as powerful tools for \textit{right now}, but our Study 2 models show that current Wikipedia editors are still slightly less likely to edit underrepresented topic articles than their baseline recommendations. In the long term, therefore, we would expect the benefits of more equity-oriented recommendations to top out and not be as useful for the underrepresented gaps that are even more distant in relevance---i.e. less familiar---to the preferences of the existing editor base.

Further, recommender systems do not help address Wikipedia's more subtle equity issues. \citeauthor{menking_wpnot_2021} (\citeyear{menking_wpnot_2021}) for example, highlight how Wikipedia's demographics inform the encyclopedia's core epistemic practices, making it difficult for those within the community to make more than incremental changes in countering systemic biases. As with any online community, these deeper issues can only be addressed through the ongoing efforts to diversify its membership.

\section{Limitations}

We now outline our work's main limitations.

First, we only explored three dimensions of content equity (gender, geography, and important topics), and even the dimensions we did study were relatively high-level. For example, while we saw an increase in global south editing during our study period, about 20\% of those edited recommendations pertained to a single country: India. On the gender side, \textit{all} of the edited biography recommendations were about cisgender individuals; 56 recommended biographies about transgender and non-binary people prompted no edits. This was unsurprising given 1.5\% uptake, but highlights that specific attention must be given to biographies of transgender and non-binary people if any measurable progress is to be made on improving them. Content equity on Wikipedia must be achieved across a multitude of dimensions, granularities, and their intersections that we do not study here. However, the consistency of our results along the dimensions we \textit{did} study gives us reason to be optimistic about how our findings generalize to the ones we did not study when they are given specific attention. 

We also only studied English Wikipedia. While we suggest ways in which our findings could generalize to other peer production platforms, more study in other contexts is needed to confidently assert generalizability. 

Similarly, we conducted our experiment on a single recommender system, launched in 2007. Had we run our experiment on a more modern recommender with substantially more accurate personalization, it is possible we would have seen a greater drop in editing when giving editors less relevant items. Our offline analysis suggests this is not the case---Wikipedians preferred editing biographies of women in the original highest-relevance recommendation sets too---but further study is needed to confidently generalize to other systems.

\section{Conclusion}

In two empirical studies on Wikipedia's SuggestBot, we found that editors were slightly \textit{more} likely to edit articles from underrepresented categories as long as item relevance was held constant. This suggests recommender systems could be used to improve content gaps on Wikipedia and other peer production platforms. From this, we discussed how recommender systems on peer production platforms might unnecessarily narrow the content they provide users when the metrics from which they make inferences do not capture information about how the user \textit{discovers} content. This paper demonstrates recommender systems' potential in improving content gaps, but we also acknowledge that such systems are only one part of a larger set of solutions for making online knowledge more equitable.

\section{Broader Perspective, Ethics and Competing Interests}
In this work, we conducted a live experiment on a recommender system for editors on English Wikipedia. The English Wikipedia editor community is very clear that Wikipedia should not be treated as a laboratory and care should be taken to avoid being disruptive.\footnote{\url{https://en.wikipedia.org/wiki/Wikipedia:What_Wikipedia_is_not#Wikipedia_is_not_a_laboratory}} As such, we took several steps to ensure that our research was not disruptive for that community:
\begin{itemize}
    \item {Consent and opt-out: the experimental recommendations were accompanied by a statement indicating that we were running an experiment, linked to a form for opting out (1 request), and monitored SuggestBot's discussion page for questions or concerns (there were none about the experiment).}
    \item {Minimize potential for harm: we designed the experiment to carefully balance need for statistical power with minimally disrupting the user experience (if the intervention turned out to be of negative impact). We left half of the recommendations unchanged and used lower-relevance but still ``reasonable'' recommendations for the interventions. We were also only providing recommendations and not editing Wikipedia itself.}
    \item {Review and approval: while English Wikipedia does not have a formal process for reviewing proposed research, we did receive IRB approval from the first author's institution and incorporated extensive feedback from researchers who have previously run experiments with SuggestBot.}
    \item {Public data: while we had access to SuggestBot to make the algorithmic changes, all recommendations and edits are public data, so we were not handling any private information.}
\end{itemize}

Implementing changes to recommender systems on Wikipedia should only be taken in careful consultation with the affected editor communities as discussed by~\citeauthor{johnson2022considerations} (\citeyear{johnson2022considerations}) (something we could not do to preserve the integrity of the experiment) and should probably not be hidden algorithmic tweaks but controllable end-user options (for example:~\citeauthor{ekstrand2015letting}~\citeyear{ekstrand2015letting}). Otherwise, the interventions could back-fire if they are irrelevant or clash with editor motivation for editing, as discussed in the Introduction.

\section{Acknowledgments}

We owe many thanks to SuggestBot's users for being part of this work. We are also grateful to the Wikidata community, which contributed much of the structured data that enabled our analyses.

\bibliography{aaai22}

\end{document}